\documentclass[10pt, sigplan, nonacm, screen]{acmart}
\usepackage{iftex}
\ifLuaTeX
  \usepackage{fontsetup}
  \let\mathscr\mathcal
\else
  \usepackage[T1]{fontenc}
  \usepackage[utf8]{inputenc}
  \usepackage{amsfonts}
  \usepackage{mathrsfs}\let\mathcal\mathscr
\fi

\usepackage{microtype}

\definecolor{Burgundy}{RGB}{128,0,32}
\usepackage{enumerate}
\usepackage{amsthm, amsmath}
\usepackage{tikz,tikz-cd}
\tikzset{commutative diagrams/arrow style=math font}

\usepackage{soul}
\usepackage[english]{babel}
\usepackage{fontawesome} 
\usepackage{xpunctuate}
\usepackage{hyperref}

\newcommand{\extlink}{~\ensuremath{{}^\text{\faExternalLink}}}

\usepackage{listings}
\makeatletter
\AtBeginDocument{%
  \let\c@lstlisting\relax
  \newcounter{lstlisting}[section]
  \gdef\thelstlisting%
      {\ifnum \c@section>\z@ \thesection.\fi \@arabic\c@lstlisting}}
\makeatother

\definecolor{keywordcolor}{rgb}{0.7, 0.1, 0.1}   
  \definecolor{keywordcolor}{cmyk}{0,0.90,0.86,0} 
\definecolor{tacticcolor}{rgb}{0.0, 0.1, 0.6}    
  \definecolor{tacticcolor}{cmyk}{1,0.1,0,0.1} 
\definecolor{commentcolor}{rgb}{0.4, 0.4, 0.4}   
  \definecolor{commentcolor}{cmyk}{1,0.1,0,0.1} 
\definecolor{symbolcolor}{rgb}{0.0, 0.1, 0.6}    
  \definecolor{symbolcolor}{cmyk}{1,0.1,0,0.1} 
\definecolor{sortcolor}{rgb}{0.1, 0.5, 0.1}      
  \definecolor{sortcolor}{cmyk}{0.20,0,1,0.19}    
\definecolor{attributecolor}{rgb}{0.7, 0.1, 0.1} 
  \definecolor{attributecolor}{cmyk}{0,0.90,0.86,0} 

\newcommand{\code}[1]{\lstinline{#1}}

\newcommand*{\mathlib}{\texttt{Mathlib}\xspace}
\theoremstyle{plain}
\newtheorem{theorem}{Theorem}[section]
\newtheorem{proposition}[theorem]{Proposition}

\theoremstyle{definition}
\newtheorem{definition}[theorem]{Definition}

\theoremstyle{remark}

\DeclareMathOperator{\Frac}{Frac} 

\AtBeginDocument{%
  }

\begin{document}

\title[Formalizing Polynomial Laws and the Universal Divided Power Algebra]{Formalizing Polynomial Laws \\
and the Universal Divided Power Algebra}

\title{Formalizing Polynomial Laws and the Universal Divided Power Algebra}

\author{Antoine Chambert-Loir}
\authornote{Both authors contributed equally to this research.}
\orcid{0000-0001-8485-7711}
\affiliation{%
	\institution{Université Paris Cité}
	\city{Paris}
	\country{France}
}
\email{antoine.chambert-loir@u-paris.fr}

\author{María Inés de Frutos-Fernández}
\authornotemark[1]
\orcid{0000-0002-5085-7446}
\affiliation{%
	\institution{Universit\"at Bonn}
	\city{Bonn}
	\country{Germany}
}
\email{midff@math.uni-bonn.de}

\renewcommand{\shortauthors}{A.~Chambert-Loir and M.~I.~de Frutos-Fernández}

\begin{abstract}
The goal of this paper is to present an ongoing formalization, in the framework provided by the Lean/Mathlib mathematical library, of the construction by Roby (1965) of the universal divided power algebra. This is an analogue, in the theory of divided powers, of the classical algebra of polynomials. It is a crucial tool in the development of crystalline cohomology; it is also used in $p$-adic Hodge theory to define the crystalline period ring. 
As an algebra, this universal divided power algebra has a fairly simple definition that shows that it is a graded algebra. The main difficulty in Roby's theorem lies in constructing a divided power structure on its augmentation ideal. To that aim, Roby identified the graded pieces with another universal structure: homogeneous polynomial laws.

We formalize the first steps of the theory of polynomial laws and show how future work will allow to complete the formalization of the above-mentioned divided power structure.
We report on various difficulties that appeared in this formalization: taking care of universes, extending to semirings some aspects of the Mathlib library, and coping with several instances of “invisible mathematics”.
\end{abstract}




\maketitle

\def\Id{\mathrm{id}}
\def\N{\mathbb N}
\def\Q{\mathbb Q}
\def\R{\mathbb R}
\def\Z{\mathbb Z}
\def\crys{{\text{\upshape crys}}}
\def\coeff{\operatorname{coeff}}
\def\Range{\operatorname{range}}

\section{Introduction}\label{sec:intro}
\subsection{Universal Properties}\label{subsec:universal}

In many areas of mathematics, including arithmetic geometry, it is common practice to work with objects defined in terms of \textit{universal properties}. That means that, instead of providing a direct definition of an object, one characterizes it by the properties it must satisfy, in such a way that any other object satisfying the same properties must be isomorphic to the first\footnote{Essentially, claiming that an object satisfies a universal property is equivalent to this object being a universal object of some auxiliary category. We refer to \cite[Chapter 2]{Riehl17} for a more detailed introduction to universal properties. }. 

A simple example of ``universal'' object would be the product of two sets $X$ and $Y$, which is the set $X \times Y$ satisfying the universal property that, for any set $Z$, maps $Z \to X \times Y$ are in bijection with pairs of maps ($Z \to X, Z \to Y$), in a compatible way with precomposition by maps of sets $W \to Z$.
Another example is that of a ring $R$ given by generators $(x_1,\dots,x_n)$ and relations $f_1,\dots,f_m \in \Z[x_1,\dots,x_n]$;
then morphisms $f\colon R \to S$ correspond to families $(s_1,\dots,s_n)$ in~$S$ satisfying 
$f_1(s_1,\dots,s_n)=\dots=f_m(s_1,\dots,s_n)=0$.

A third example, more deeply relevant to the present paper, is the algebra
of polynomials over a given ring~$R$, which is the universal
commutative and associative $R$-algebra over a set of generators;
this construction can be extended to that of the symmetric algebra of an
arbitrary $R$-module~$M$, which is the universal $R$-algebra~$S_R(M)$ endowed
with an $R$-linear map $M\to S_R(M)$. This algebra has actually more
structure: similarly to the fact that a polynomial can be uniquely
decomposed as a sum of its homogeneous components, the $R$-algebra~$S_R(M)$
is a direct sum of components $S^d_R(M)$, for $d\in\N$, such 
that the product of elements in $S^d_R(M)$ and $S^e_R(M)$ belongs
to $S^{d+e}_R(M)$:  we say that $S_R(M)$ is a \emph{graded algebra}.
Observe that the sum $S^+_R(M)$ of all $S^d_R(M)$, for $d>0$, is an ideal
of $S_R(M)$, called its \emph{augmentation ideal.}
Classical polynomials in $n$~indeterminates
are recovered when one consider the free $R$-module of rank~$n$,
and the augmentation ideal then corresponds to the ideal of polynomials
without constant term.

Since universal objects are essentially unique, once one proves that there exists an object satisfying a given universal property, it is 
often possible to ``forget'' the concrete construction of the object and instead simply use the fact that it satisfies the desired property. Then, the universal properties can be viewed as computational rules or induction principles.
However, in order to establish some  properties of universal objects, it may be important to have
an alternative description, e.g., an explicit construction. This is the case, for example, for polynomial algebras,
which are universal commutative and associative rings on a set of generators, 
but whose explicit description using monomials is needed to show that they are free modules.

\subsection{The Universal Divided Power Algebra}\label{subsec:dpenv}
In this paper, we are particularly interested in two universal objects, the \textit{divided power envelope} and the \textit{universal divided power algebra}, that, despite their complexity, play a key role in modern arithmetic geometry.

The divided power envelope was used by Berthelot \cite{Berthelot-1974, BerthelotOgus-1978} to develop the theory of crystalline cohomology (which, together with subsequent $p$-adic cohomologies, constitutes one of the tools used in the proof of Fermat's Last Theorem by Wiles and his collaborators). It is also needed to define the Fontaine period ring $B_\crys$, which is used in $p$-adic Hodge theory to study $p$-adic Galois representations and to compare between different cohomology theories. 

The construction of the divided power envelope relies on first defining the universal divided power algebra of a module, which is our main focus in this paper. While arithmetic geometers most often treat these two objects as black boxes, if we wish to formalize results from this area in Lean, we must first formalize these constructions. Although a 15-page summary of these constructions is provided in an appendix to \cite{BerthelotOgus-1978}, the full details require over 100 pages of mathematical research. 
While the formalization of these definitions will therefore require significant work, it is also an important task, in particular because Roby's original construction of the universal divided power algebra \cite{Roby-1965} contained a mistake (although an alternative proof that avoids this issue has been sketched by Berthelot and Ogus in \cite{BerthelotOgus-1978}).

Given an ideal $I$ in a commutative ring $R$, a \textit{divided power structure} on $I$ is a family of maps $\{\gamma_n : I \to R \}$ indexed by the natural numbers that behave analogously to the family $x \mapsto \frac{x^n}{n!}$, even when division by factorials is not necessarily defined in $R$ (see Definition \ref{definition:dp} for more details). We then say that $I$ has \textit{divided powers}, and that $(R, I, \gamma)$ is a \textit{divided power algebra}.
For example, if $R = \Z_p$ is the ring of $p$-adic integers, then the family $x \mapsto \frac{x^n}{n!}: (p) \to \Z_p$ is a divided power structure on the ideal $(p)$, despite $p$ not being an invertible element in~$\Z_p$. This follows from the (non-obvious) fact that for every natural number~$n$, the exponent of~$p$ in
the decomposition of~$n!$ into prime powers is at most~$n$.
There are notions of \textit{divided power morphisms} between divided power algebras, as well as of \textit{sub-divided power ideals}. A nice introduction to the theory is given in \cite[Chapter~3]{BerthelotOgus-1978}.

The \textit{universal divided power algebra} of an $R$-module~$M$ is a triple $(\Gamma_R(M), \Gamma_R^+(M), \gamma_M)$, where $\Gamma_R(M)$ is a graded algebra together with an $R$-linear map $\varphi \colon M \to \Gamma_R(M)$, $\Gamma_R^+(M)$ is the augmentation ideal of $\Gamma_R(M)$, and $\gamma_M$ is a divided power structure on $\Gamma_R^+(M)$, such that given a divided power algebra $(S, J, \gamma_S)$ and an $R$-linear map $\psi \colon M \to J$, there is a unique divided power morphism $\tilde{\psi} \colon (\Gamma_R(M), \Gamma_R^+(M), \gamma_M) \to (S, J, \gamma_S)$ such that $\tilde{\psi} \circ \varphi = \psi$.

At this point, it is probably useful to give the simplest
example of the structure that one obtains,  considering
the case of a free module of rank~$1$. Then one can prove
that $\Gamma_R(M)$ is the set of all finite sums
$\sum r_n T^{(n)}$, with $r_n\in R$ for all~$n$,
 where $T^{(n)}$ is a formal symbol, subject to the obvious
addition, and to the multiplication given by 
$T^{(m)}T^{(n)}= \binom {m+n}n T^{(m+n)}$. 
When $R$ is a subring of a field~$F$ of characteristic zero,
we can view this algebra as a subring of the algebra 
of polynomials $F[T]$ with coefficients in~$F$,
by interpreting~$T^{(n)}$ as $T^n/n!$.

Defining the ring $\Gamma_R(M)$ is easy, and the proof that it is a graded ring is relatively straightforward. However, the construction of the divided power structure $\gamma_M$ is a hard theorem of Roby \cite{Roby-1963,Roby-1968}. In this article, we present an ongoing formalization of this proof.
A key step in Roby's construction is the identification of the graded components of $\Gamma_R(M)$ with certain spaces of \textit{polynomial laws}. Hence, as a prerequisite, we formalize the theory of polynomial laws. 

Polynomial laws can be thought of as generalizations of maps $\R^n \to \R$ determined by multivariate polynomials (see Definition \ref{definition:pl}). As in the classical setting of polynomials, there are notions of coefficients, homogeneous polynomial laws, differentials and partial derivatives. We formalize these definitions in Lean, and provide a significant API to work with them.

Besides their important role in Roby's construction of the universal divided power algebra, polynomial laws have been used by Chenevier to study rigid analytic families of Galois representations \cite{Chenevier14}. 

To the authors knowledge, this paper describes the first formalization of the theory of polynomial laws in any theorem prover. Our ongoing formalization of the universal divided power algebra builds over our formalization (in Lean~4) of the basic theory of divided powers \cite{CLdFF25}. This theory has not yet been formalized in any other theorem prover.

\subsection{Lean and Mathlib}\label{subsec:lean} 
This formalization project is written in Lean~4 \cite{Lean4} and builds over Lean's mathematical library \mathlib \cite{Mathlib}. It includes around 10,500 new lines of code, most of which can be
found in the public GitHub repository \url{https://github.com/mariainesdff/divided_power_algebra_journal}. The remaining of this formalization work, including the definition of polynomial laws, has already been merged into Mathlib.

Some of the code snippets in the article have been lightly edited for readability (for example, we might omit comments, namespaces, or details about variables being implicit). Throughout the text, we use the clickable symbol {\faExternalLink} to link to the source code for the
formalization.

Note that many of our definitions are marked as noncomputable. This is currently unavoidable, since we are building over definitions which are noncomputable in \mathlib (for instance, \mathlib's polynomials are noncomputable).

\subsection{Paper Outline}\label{subsec:outline}

In \S\ref{sec:poly}, we describe the formalization of the theory of polynomial laws: after motivating the definition in \S\ref{subsec:poly-def}, we introduce coefficients in \S\ref{subsec:coeff}, homogeneity in \S\ref{subsec:homog} and differentials and partial derivatives in \S\ref{subsec:diff}. \S\ref{sec:dpa} is devoted to the universal divided power algebra: \S\ref{subsec:dp} reviews the definition of divided powers, \S\ref{subsec:udpa} introduces the universal divided power algebra, \S\ref{subsec:dp-gr} its graded structure and \S\ref{subsec:dpow-const} its divided power structure, while \S\ref{subsec:basech} contains a base change result.
In \S\ref{sec:impl}, we present a detailed discussion of the main difficulties encountered during the formalization process, related to instances (\S\ref{subsec:instances}), universes (\S\ref{subsec:univ}), extension of results to semirings (\S\ref{subsec:semir}), ``invisible mathematics'' (\S\ref{subsec:invis}), and automation (\S\ref{subsec:auto}).
Finally, we summarize our conclusions in \S\ref{sec:concl}.

\section {Polynomial Laws}\label{sec:poly}

\subsection{Definition}\label{subsec:poly-def}

To give motivation for the definition of a polynomial law,
let us start with the elementary example of a polynomial map
from $\R^2$ to $\R$: nothing but a map $f\colon\R^2\to\R$
for which there exists a polynomial $p\in\R[X,Y]$
such that $f(x,y)=p(x,y)$ for every $(x,y)\in\R^2$. 
Here, $p$ is a “formal” finite sum $\sum c_{m,n}X^m Y^n$,
and $X,Y$ are two indeterminates, the $(c_{m,n})$ being the
coefficients of~$p$.
Because the field of real numbers is infinite, one can prove
that the polynomial~$p$ is uniquely determined by the condition
that it represents the function~$f$.
On the other hand, the existence of~$p$ allows one to \emph{extend}
the map~$f$ to other contexts. In particular, if $(x,y)$
is a pair of complex numbers, one can define $f(x,y)$
by the formula $f(x,y)=\sum c_{m,n} x^m y^n$.
Denoting complex conjugation by~$\overline{\cdot}$,
this algebraic expression guarantees that the formula
$\overline{f(x,y)}=f(\overline x,\overline y)$ is satisfied.
The entries~$x$ and~$y$ can also be assumed to belong to 
any other commutative $\R$-algebra, with similar compatibilities
with respect to morphisms of algebras.

The mathematical definition of a \emph{polynomial law} generalizes
this example in two directions:
\begin{enumerate}
\item One replaces the field~$\R$
of real numbers by any commutative ring~$R$,
\item One replaces the vector spaces $\R^2$ and~$\R$ with arbitrary
$R$-modules~$M$ and~$N$,
\end{enumerate}
leading to the definition below, which 
was apparently first introduced by Roby in~\cite[I.2]{Roby-1963}.

Before we give the definition, recall that one can construct,
associated to every $R$-module~$M$ and every $R$-algebra~$S$,
an $S$-module~$M_S$, 
in a universal way,
so that any $R$-linear map $f\colon M\to N$ between two $R$-modules~$M$
and~$N$
extends naturally to an $S$-linear map $f_S\colon M_S\to N_S$,
compatible with composition and addition on linear maps.
In fact, $M_S$ is defined by the tensor product $S\otimes_R M$
and the map $f_S$ associates with a split tensor $s\otimes m$
in $S\otimes_R M$ the element $s \otimes f(m)$ of $S\otimes_R N$.
Roby's intuition is that for “polynomial laws”, this extension 
principle still should hold by definition.

\begin{definition}\label{definition:pl}
Let $R$  be a commutative ring and let~$M$, $N$ be $R$-modules.
An \emph{$R$-polynomial law}~$f$ from~$M$ to~$N$
is the datum, for every commutative $R$-algebra~$S$,
of a map 
\[ f_S \colon S\otimes_R M \to S\otimes_R N\] subject to the
condition
\[ (\phi \otimes \Id_N) \circ f_S = f_{S'} \circ (\phi\otimes \Id_M), \]
for every morphism $\phi\colon S\to S'$ of commutative $R$-algebras.
\end{definition}

The \lstinline{structure} \lstinline{PolynomialLaw}\href{https://github.com/leanprover-community/mathlib4/blob/91e4c39b71414a092707d1299388397e761f8dd7/Mathlib/RingTheory/PolynomialLaw/Basic.lean#L58}{\extlink}
in \mathlib formalizes this mathematical definition.
Its field \lstinline{PolynomialLaw.toFun'}
contains the datum of the maps~$f_S$, and
the functoriality property is reflected by
the field \lstinline{PolynomialLaw.isCompat'}.
Two polynomial laws~$f$ and~$g$ are equal if and only if the
maps~$f_S$ and~$g_S$ are equal, for all~$S$; this is reflected by
the \lstinline{ext} attribute to the structure \lstinline{PolynomialLaw}.

\begin{lstlisting}[caption={Definition of a polynomial law.},float=htb, label=code:pl,captionpos=t,abovecaptionskip=-\medskipamount]
@[ext]
structure PolynomialLaw (R : Type u) [CommSemiring R]
    (M N : Type*) [AddCommMonoid M] [Module R M] [AddCommMonoid N] [Module R N] where
 toFun' (S : Type u) [CommSemiring S] [Algebra R S] : 
   S ⊗[R] M → S ⊗[R] N
 isCompat' {S : Type u} [CommSemiring S]
   [Algebra R S] {S' : Type u} [CommSemiring S']
   [Algebra R S'] (φ : S →ₐ[R] S') :
   φ.toLinearMap.rTensor N ∘ toFun' S = 
     toFun' S' ∘ φ.toLinearMap.rTensor M := by aesop
\end{lstlisting}

We write $\mathscr P_R(M;N)$ for the set of $R$-polynomial laws from~$M$
to~$N$, denoted by \lstinline{M →ₚₗ[R] N} in \mathlib.
Its natural structure of an $R$-module 
has been formalized in \mathlib as an instance\href{https://github.com/leanprover-community/mathlib4/blob/91e4c39b71414a092707d1299388397e761f8dd7/Mathlib/RingTheory/PolynomialLaw/Basic.lean#L153}{\extlink}.

\subsection{Coefficients}\label{subsec:coeff}

While Definition~\ref{definition:pl}
does not make it obvious that these objects should be 
thought of as polynomials, polynomial laws have coefficients once one
instantiates a finite family $(m_i)_{i\in  I}$ of  vectors in~$M$ as follows.

Let $f\in\mathscr P_R(M;N)$.
Consider the ring $R[T]:=R[(T_i)]$ of polynomials with coefficients
in~$R$ in indeterminates~$T_i$, for $i\in I$, and the
element $m_T = \sum_{i\in I} T_i\otimes m_i$ of $R[T]\otimes_R M$.
The element $f_{R[T]}(m_T)$ belongs to $R[T]\otimes_R N$, hence
can be uniquely written as as finite sum $\sum n_a T^a$,
for a family $(n_a)$ of elements of~$\N$ indexed by all $a\colon \N^{(I)}\to \N$. These are the \emph{coefficients} of~$f$ relative to the family~$(m_i)$.
We write $n_a = \coeff(f,m)_a$.

Relevant parts of the Lean definition are given in code excerpt~\ref{code:pl-coeff}. The expression $\sum_{i \in I},  T_i \otimes m_i$ is coded in Lean
via a definition \lstinline{Module.generize}\href{https://github.com/mariainesdff/divided_power_algebra_journal/blob/beb12bf28df8bad96ed075e3f3dc99b191bfee0c/DividedPowerAlgebra/PolynomialLaw/Coeff.lean#L50}{\extlink} for which some API is provided. 

We then define the associated \code{PolynomialLaw.generize'}\href{https://github.com/mariainesdff/divided_power_algebra_journal/blob/beb12bf28df8bad96ed075e3f3dc99b191bfee0c/DividedPowerAlgebra/PolynomialLaw/Coeff.lean#L135}{\extlink} as the linear map sending a polynomial law $f \in \mathscr P_R(M;N)$ to
$f (\sum_{i \in I},  T_i \otimes m_i) \in R[T]\otimes_R N$, as an auxiliary step for the definition of \code{PolynomialLaw.coeff}\href{https://github.com/mariainesdff/divided_power_algebra_journal/blob/beb12bf28df8bad96ed075e3f3dc99b191bfee0c/DividedPowerAlgebra/PolynomialLaw/Coeff.lean#L142}{\extlink}, which inherits this linearity.

\begin{lstlisting}[caption={Coefficients of a polynomial law.},float=htb, label=code:pl-coeff,captionpos=t,abovecaptionskip=-\medskipamount]
variable {ι : Type*} [Fintype ι] [DecidableEq ι]

noncomputable def generize :
    (ι → M) →ₗ[R] MvPolynomial ι R ⊗[R] M where
  toFun m := ∑ i, X i ⊗ₜ[R] m i
  ...

noncomputable def generize' (m : ι → M) :
    (M →ₚₗ[R] N) →ₗ[R] MvPolynomial ι R ⊗[R] N where
  toFun f := f.toFun (MvPolynomial ι R) (Module.generize m)
  ...

noncomputable def PolynomialLaw.coeff :
    (M →ₚₗ[R] N) →ₗ[R] (ι →₀ ℕ) →₀ N :=
  scalarRTensor.toLinearMap.comp (generize m)
\end{lstlisting}

The structure of the $R$-module $\mathscr P_{R}(M;N)$
of polynomial laws is very complicated in general. However, it simplifies a lot when the $R$-module~$M$ is free of finite rank.

\begin{proposition}
Let $M$ be a free $R$-module of finite rank with $R$-basis $m=(m_i)_{i\in I}$.
Then the map $f \mapsto \sum_a \coeff(f;m)_a T^a$ 
gives an $R$-linear isomorphism  from $\mathscr P_R(M;N)$
to the module $N[(T_i)_{i\in I}]$
of polynomials of coefficients in~$N$ in indeterminates~$T_i$.
\end{proposition}

This isomorphism (in fact, its inverse) is provided as a
\lstinline{LinearEquiv} called \code{polynomialLawEquivCoeff}\href {https://github.com/mariainesdff/divided_power_algebra_journal/blob/beb12bf28df8bad96ed075e3f3dc99b191bfee0c/DividedPowerAlgebra/PolynomialLaw/Coeff.lean#L438}{\extlink}.

\subsection{Homogeneous Polynomial Laws}\label{subsec:homog}

\begin{definition}
Let $d$ be a natural number.
A polynomial law $f\in\mathscr P_R(M;N)$ is said to be homogeneous
of degree~$d$ if the relation
\[ f_S(s m) = s^d f_S(m) \]
holds for any commutative $R$-algebra~$S$,
any $s\in S$ and any $m\in S\otimes_R M$.
\end{definition}

We formalize this definition as a \lstinline{Prop}-valued
function \lstinline{PolynomialLaw.IsHomogeneous}\href{https://github.com/mariainesdff/divided_power_algebra_journal/blob/beb12bf28df8bad96ed075e3f3dc99b191bfee0c/DividedPowerAlgebra/PolynomialLaw/Homogeneous.lean#L80}{\extlink}.
Homogeneous polynomial laws of degree~$d$ form a submodule
$\mathscr P_R(M;N)_d$ of $\mathscr P_R(M;N)$,
formalized as definition \lstinline{PolynomialLaw.grade}\href{https://github.com/mariainesdff/divided_power_algebra_journal/blob/beb12bf28df8bad96ed075e3f3dc99b191bfee0c/DividedPowerAlgebra/PolynomialLaw/Homogeneous.lean#L94}{\extlink} in Lean.
 
\begin{lstlisting}[caption={Homogeneous polynomial laws.},float=htb, label=code:homog-pl,captionpos=t,abovecaptionskip=-\medskipamount]
def IsHomogeneous (d : ℕ) (f : M →ₚₗ[R] N) : Prop :=
  ∀ (S : Type u) [CommSemiring S] [Algebra R S]
    (r : S) (m : S ⊗[R] M),
    f.toFun' S (r • m) = r ^ d • f.toFun' S m

def grade (d : ℕ) : Submodule R (M →ₚₗ[R] N) where
  carrier := IsHomogeneous d
  ...
\end{lstlisting}
Analogously to the decomposition of a polynomial into its homogeneous
components, any polynomial law~$f\in\mathscr P_R(M;N)$ 
has homogeneous components~$f_d$, for $d\in\N$,
so that for every commutative $R$-algebra~$S$ and any $m\in S\otimes_R M$,
one has 
\begin{equation}\label{eq:decomp}
f_S(m) = \sum_{d\in\N} f_{d,S}(m).
\end{equation} 
To \emph{define}~$f_d$, one uses the expected compatibility properties:
introducing an indeterminate~$T$ and
considering  $T \otimes m \in S[T]\otimes_R M$, one should have

\[ f_{S[T]}(T \otimes m)=\sum_{d} f_{d,S[T]}(T \otimes m)
 = \sum_{d} T^d \otimes f_{d,S} (m), \]
which leads to the definition of~$f_{d,S}$ as the coefficient
of degree~$d$ of~$f_{S[T]} (T \otimes m)\in S[T]\otimes_R N$ viewed
as an element of the polynomial module $(S\otimes_R N)[T]$.
The fact that~$f_d$ is actually a polynomial law is deduced
by the compatibility condition of the polynomial law~$f$ in the diagram
of $R$-algebras
\[ \begin{tikzcd} S \ar{r}{\phi} \ar{d} & S' \ar{d} \\
   S[T] \ar{r}{\phi_T}  & S'[T] \end{tikzcd}, \]
where $\phi\colon S\to S'$ is a morphism of commutative $R$-algebras
and $\phi_T\colon S[T]\to S'[T]$ is the corresponding morphism
of polynomials algebras deduced from applying~$\phi$ on coefficients.

We provide  \code{PolynomialLaw.component}\href{https://github.com/mariainesdff/divided_power_algebra_journal/blob/beb12bf28df8bad96ed075e3f3dc99b191bfee0c/DividedPowerAlgebra/PolynomialLaw/Homogeneous.lean#L423}{\extlink} 
for the definition of the component $f_d$, which we show to be homogeneous in \code{component_isHomogeneous}\href{https://github.com/mariainesdff/divided_power_algebra_journal/blob/beb12bf28df8bad96ed075e3f3dc99b191bfee0c/DividedPowerAlgebra/PolynomialLaw/Homogeneous.lean#L450}{\extlink}, and we prove the decomposition formula (\ref{eq:decomp}) in \code{lfsum_component}.\href{https://github.com/mariainesdff/divided_power_algebra_journal/blob/beb12bf28df8bad96ed075e3f3dc99b191bfee0c/DividedPowerAlgebra/PolynomialLaw/Homogeneous.lean#L514}{\extlink}

Homogeneous polynomial laws of degrees~$0$, $1$ and~$2$
correspond to classical objects: constant maps, linear maps,
and (a general notion of) quadratic maps.
The case of linear maps is particularly important:
by base change, a linear map $f\in\operatorname{Hom}_R(M;N)$
gives rise to linear maps $f_S \colon S\otimes_R M\to S\otimes_R N$,
(known to \mathlib as \lstinline{LinearMap.baseChange}\href{https://github.com/leanprover-community/mathlib4/blob/91e4c39b71414a092707d1299388397e761f8dd7/Mathlib/LinearAlgebra/TensorProduct/Tower.lean#L599}{\extlink}),
for all $R$-algebras~$S$, and this family $(f_S)$ is a polynomial law\href{https://github.com/mariainesdff/divided_power_algebra_journal/blob/beb12bf28df8bad96ed075e3f3dc99b191bfee0c/DividedPowerAlgebra/PolynomialLaw/Basic.lean#L521}{\extlink}.
This construction gives rise to an $R$-linear isomorphism
from $\operatorname{Hom}_R(M;N)$ to $\mathscr P_R(M;N)_1$,
which we formalized as \code{LinearMap.toDegreeOnePolynomialLawEquiv}\href{https://github.com/mariainesdff/divided_power_algebra_journal/blob/beb12bf28df8bad96ed075e3f3dc99b191bfee0c/DividedPowerAlgebra/PolynomialLaw/Homogeneous.lean#L381}{\extlink}.

Compared to the case of polynomials, which  only have finitely many
nonzero homogeneous components,
a polynomial law~$f$ 
can have nonzero homogeneous components~$(f_d)$ of all degrees when the module~$M$ is not finitely generated.
For instance, if $M$ is an $\R$-vector space with a countable basis $(b_n)_{n \in \N}$, then the function sending $x = \sum_n x_n b_n \in M$ to $\sum_n x_n^n \in \R$ induces a polynomial law $f \in P_R(M;\mathbb{R})$ whose $n^\text{th}$ homogeneous component is induced by the function $x \mapsto x_n^n$; hence all homogeneous components of $f$ are nonzero \cite[\S IV.6]{Roby-1963}.
This example shows that the module $\mathscr P_R(M;N)$ may lie
strictly between the direct sum $\bigoplus_d \mathscr P_R(M;N)_d$
and the direct product $\prod_d \mathscr P_R(M;N)_d$. 

However, for every~$S$ and every $m\in S\otimes_R M$,
the family $(f_{d,S}(m))$ has only finite support.
We formalize this property by saying that the family $(f_d)$
is \emph{locally finite} (which we denote \code{PolynomialLaw.LocFinsupp}\href{https://github.com/mariainesdff/divided_power_algebra_journal/blob/5d1dafcd4c59b7de9791416bfc47442e6b5d2324/DividedPowerAlgebra/PolynomialLaw/LocFinsupp.lean#L38}{\extlink} in Lean) and provide an API for the summation
of locally finite families of polynomial laws\href{https://github.com/mariainesdff/divided_power_algebra_journal/blob/5d1dafcd4c59b7de9791416bfc47442e6b5d2324/DividedPowerAlgebra/PolynomialLaw/LocFinsupp.lean}{\extlink}
(eventually extended by~$0$ when the family is not locally finite).

\begin{lstlisting}[caption={Locally finite families of polynomial laws and their sums.},float=htb, label=code:lfsum,captionpos=t,abovecaptionskip=-\medskipamount]
variable {ι : Type*} (f : ι → M →ₚₗ[R] N})

def LocFinsupp : Prop :=
  ∀ (S : Type u) [CommSemiring S] [Algebra R S]
    (m : S ⊗[R] M),
    (fun i ↦ (f i).toFun' S m).support.Finite

noncomputable def LocFinsupp.sum
    (hf : LocFinsupp f) : M →ₚₗ[R] N where
  toFun'    := fun S _ _ m ↦
    (ofSupportFinite _ (hf S m)).sum fun _ m ↦ m
  isCompat' := fun {S _ _ S' _ _} φ ↦ by ...

noncomputable def lfsum : M →ₚₗ[R] N :=
  if hf : LocFinsupp f then hf.sum else 0
\end{lstlisting}

In the case where $M$ is a product module $M_1\times\dots \times M_p$,
Roby develops in~\cite[\S~I.8]{Roby-1963} a variant of the notion of homogeneity 
that considers the degrees on each factor, and uses that variant for the definition of a differential calculus
on polynomial laws. Therefore, we formalized this notion in definition \code{IsMultiHomogeneous}\href{https://github.com/mariainesdff/divided_power_algebra_journal/blob/5d1dafcd4c59b7de9791416bfc47442e6b5d2324/DividedPowerAlgebra/PolynomialLaw/MultiHomogeneous.lean#L67}{\extlink}, for which we provided an API analogous to that described in this section. To facilitate its study, we also formalized 
a specialization of the definition of coefficient to this setting, denoted \code{multiCoeff}\href{https://github.com/mariainesdff/divided_power_algebra_journal/blob/5d1dafcd4c59b7de9791416bfc47442e6b5d2324/DividedPowerAlgebra/PolynomialLaw/MultiCoeff.lean#L137}{\extlink}. While mathematically these two notions of homogeneity are very similar, the formalization of the latter presents some subtleties due to implicit isomorphisms between tensor products; see Section \ref{subsec:invis}.

Products of two objects are often treated separately in Lean (using the type \code{Prod}\href{https://github.com/leanprover/lean4/blob/ad1a017949674a947f0d6794cbf7130d642c6530/src/Init/Prelude.lean#L502}{\extlink}rather than a general product type). Following this philosophy, we provide \code{biCoeff}\href{https://github.com/mariainesdff/divided_power_algebra_journal/blob/5d1dafcd4c59b7de9791416bfc47442e6b5d2324/DividedPowerAlgebra/PolynomialLaw/BiCoeff.lean#L154}{\extlink} and \code{IsBiHomogeneous}\href{https://github.com/mariainesdff/divided_power_algebra_journal/blob/5d1dafcd4c59b7de9791416bfc47442e6b5d2324/DividedPowerAlgebra/PolynomialLaw/BiHomogeneous.lean#L60}{\extlink} for the case where $M \simeq M_1 \times M_2$.

\subsection{Differentials and Partial Derivatives}\label{subsec:diff}

\begin{definition}\label{definition:diff}(\cite[\S~II.4]{Roby-1963}]).
Let $f \in \mathscr P_R(M;N)$ be a polynomial law and let $n \in \N$. The \emph{divided differential of order~$n$ of~$f$} is the polynomial law $D^nf \in \mathscr P_R(M^2;N)$ defined by the formula
\begin{equation}\label{eq:diff}
	D^nf = \sum_{p=0}^{\infty}\Pi^{p, n}f,
\end{equation}
where $\Pi^{p, n}f$ denotes the bihomogeneous component of degree $(p, n)$ of the polynomial law $\Pi_2f \mathscr P_R(M^2;N)$ obtained by composing $f$ with the law induced by $(m_1, m_2) \mapsto m_1  + m_2$.
\end{definition}

Equivalently, for any $R$-algebra $S$ and any $(z, z') \in S \otimes_R M$,   $D^nf(z, z')$ is the sum of the coefficients of the monomials of degree $n$ in $T_2$ in the expression
\begin{equation}\label{eq:diff'}
f_{S[T_1, T_2]} (z \otimes T_1 + z' \otimes T_2). 
\end{equation}

Roby establishes several basic results about the differential of a polynomial law $f$, such as the fact that it is linear on the second term, or that if $f$ is homogeneous of degree $p$, then its $n$ divided differential vanishes for $n > p$. He also proves a \emph{Taylor formula for polynomial laws} \cite[Prop. II.2]{Roby-1963}, analogous to the classical Taylor formula for polynomials, which is a key ingredient in the construction of the divided power structure on the universal divided power algebra. This Taylor formula involves a notion of \emph{divided partial derivative} \cite[\S~II.5]{Roby-1963}. 

We have some work in progress towards formalizing these results, using both the structured definition (\ref{eq:diff}) of the divided differential\href{https://github.com/mariainesdff/divided_power_algebra_journal/blob/5d1dafcd4c59b7de9791416bfc47442e6b5d2324/DividedPowerAlgebra/PolynomialLaw/Differential.lean#L40}{\extlink}, and the one provided by the explicit formula~(\ref{eq:diff'})\href{https://github.com/mariainesdff/divided_power_algebra_journal/blob/5d1dafcd4c59b7de9791416bfc47442e6b5d2324/DividedPowerAlgebra/PolynomialLaw/Differential2.lean#L109}{\extlink}. In either case, while the mathematical arguments are expressed rather clearly in Roby's article (albeit in an imprecise language), their immediate transcription in Lean is definitely tedious. It seems that an extensive API will be needed to make them as easy as the informal proof.
\section {The Universal Divided Power Algebra}\label{sec:dpa}

\subsection{Divided Powers}\label{subsec:dp}
The notion of divided powers on a ring~$R$ was 
invented by Cartan~\cite{Cartan-1955}, who observed 
that some rings carry maps that share 
the same algebraic properties as the maps $x\mapsto x^n/n!$
on a $\Q$-algebra. 
The construction by Berthelot~\cite{Berthelot-1974}
of crystalline cohomology revealed its importance for algebraic geometry.

\begin{definition}[{\cite[Definition 3.1]{BerthelotOgus-1978}}]\label{definition:dp}
Let $I$ be an ideal in a commutative ring $R$.
A \emph{divided power structure on $I$} is a family of maps $\gamma_n \colon I \to R$ for $n \in \N$ such that the following properties hold,
for all $x,y\in I$, all $m, n\in \N$ and all $r\in R$,
\begin{enumerate}[(i)]
    \item  $\gamma_0 (x) = 1$;
    \item  $\gamma_1 (x) = x$;
    \item if $ n > 0$, then $\gamma_n (x) \in I$;
    \item $ \gamma_n (x + y) = \sum_{i + j=n} {\gamma_i(x) \cdot \gamma_j(y)}$;
    \item $ \gamma_n (r \cdot x) = r^n \cdot \gamma_n(x)$;
    \item $ \gamma_m (x) \cdot \gamma_n (x) = \binom{m + n}m \cdot \gamma_{m + n} (x)$;
    \item if $n>0$, then $ \gamma_m (\gamma_n (x)) = \frac{(m \cdot n)!}{m!(n!)^m} \cdot \gamma_{m \cdot n} (x)$.
\end{enumerate}
\end{definition}
For property~(vii), one needs to observe that the rational number
$(m\cdot n)!/m!(n!)^m$ is in fact a natural number.
Property~(iv) is a binomial formula without binomial coefficients.
From properties~(i, ii, vi), one can deduce that $x^n = n! \gamma_n(x)$ 
for all $n\in\N$ and all $x\in I$.

In prior work, we formalized this mathematical definition in Lean~\cite{CLdFF25},
as a \lstinline{structure} \lstinline{DividedPowers}\href{https://github.com/leanprover-community/mathlib4/blob/91e4c39b71414a092707d1299388397e761f8dd7/Mathlib/RingTheory/DividedPowers/Basic.lean#L74}{\extlink} together
with some elementary properties and constructions, available in \mathlib~\href{https://github.com/leanprover-community/mathlib4/tree/91e4c39b71414a092707d1299388397e761f8dd7/Mathlib/RingTheory/DividedPowers}{\extlink}.

Let $R$ be a commutative ring and let~$I$ be an ideal of~$R$. 
For general categorical reasons, there exists a commutative ring~$S$,
a ring morphism $\phi\colon R\to S$, 
an ideal~$J$ of~$S$ such that $\phi(I)\subset J$,
and a divided power structure on~$J$,
the whole being universal in some natural sense.
It can indeed be defined as a suitable
adjoint of the forgetful functor $(S,J,\gamma)\mapsto (S,J)$
that forgets the divided power structure~$\gamma$ on the ideal~$J$
of the commutative ring~$S$.
This universal triple is called the \emph{divided power envelope}
of the pair~$(R,I)$.

This construction is fundamental for the development of crystalline
cohomology. It also appears in $p$-adic Hodge theory, 
for the construction of the ring $B_{\mathrm{crys}}$
of crystalline $p$-adic periods.

However, this universal construction does not give any insight about
the properties of this divided power envelope, and a more
explicit construction is required. That construction,
due to Berthelot~\cite{Berthelot-1974}, makes use
of another explicit construction, due to Roby~\cite{Roby-1963, Roby-1965},
of the universal divided power algebra of a module.

\subsection{The Universal Divided Power Algebra}\label{subsec:udpa}

\begin{definition}\label{definition:dpa}
Let $R$ be a commutative ring and let~$M$ be an $R$-module.
The \emph{universal divided power algebra} of~$M$ is the $R$-algebra $\Gamma_R(M)$
defined by generators~$x^{[n]}$, for all $x\in M$ and all $n\in\N$,
with the following family of relations for $x,y\in M$, $n, p\in\N$ and $r\in R$:
\begin{enumerate}[(i)]
    \item $x^{[0]}= 1$;
    \item $(x+y)^{[n]} = \sum_{i + j=n} x^{[i]}\cdot y^{[j]}$;
    \item $(rx)^{[n]}= r^n x^{[n]}$;
    \item $x^{[m]}\cdot x^{[n]} = \binom{m + n}m \cdot x^{[m+n]}$.
\end{enumerate}
\end{definition}
This definition is readily formalized in Lean:
	in our code, we encode the family of relations introduced in Definition~\ref{definition:dpa} as an inductive definition \code{DividedPowerAlgebra.Rel}\href{https://github.com/mariainesdff/divided_power_algebra_journal/blob/5d1dafcd4c59b7de9791416bfc47442e6b5d2324/DividedPowerAlgebra/DPAlgebra/Init.lean#L64}{\extlink}. As an abstract ring, the \code{DividedPowerAlgebra}\href{https://github.com/mariainesdff/divided_power_algebra_journal/blob/5d1dafcd4c59b7de9791416bfc47442e6b5d2324/DividedPowerAlgebra/DPAlgebra/Init.lean#L88}{\extlink} is defined as the ring quotient of the polynomial ring in the variables $\N \times M$ with coefficients in $R$ by these relations.
 
A morphism $f\colon M\to N$ of $R$-modules gives rise
to an algebra morphism $\Gamma_R(f)\colon \Gamma_R(M)\to\Gamma_R(N)$,
given on generators by $x^{[n]}\mapsto f(x)^{[n]}$.
By these maps, the construction of the divided power algebra is functorial in the module~$M$.

The formalization of the map $(x,n)\mapsto x^{[n]}$ 
appears in a curried form as
\begin{lstlisting}
def dp (n : ℕ) (x : M) : DividedPowerAlgebra R M := 
  mkAlgHom R (Rel R M) (X ⟨n, x⟩)
\end{lstlisting}
That is, \code{dp R n m}\href{https://github.com/mariainesdff/divided_power_algebra_journal/blob/5d1dafcd4c59b7de9791416bfc47442e6b5d2324/DividedPowerAlgebra/DPAlgebra/Init.lean#L109}{\extlink} is the equivalence class of the variable~$X_{n, m}$ in \code{DividedPowerAlgebra R M}.

\paragraph{Weak universal property} The first universal property satisfied by the universal divided power algebra is the following: given an $R$-algebra $A$ with a divided power structure~$\gamma_J$ on some ideal $J \subseteq A$, together with an $R$-linear map $\varphi : M \to A$ with image contained in $J$, there exists a unique $R$-linear map $\tilde\varphi : \Gamma_R(M) \to A$ such that \[ \tilde\varphi (x^{[n]}) = \gamma_{J, n} (\varphi (x)) \quad \forall n \in \N, x \in M. \]
	
We formalize the definition of the functor $\varphi \mapsto \tilde\varphi$ as \code{DividedPowerAlgebra.lift}\href{https://github.com/mariainesdff/divided_power_algebra_journal/blob/5d1dafcd4c59b7de9791416bfc47442e6b5d2324/DividedPowerAlgebra/DPAlgebra/Init.lean#L273}{\extlink}, and prove this universal property in lemmas \code{DividedPowerAlgebra.lift_apply_dp}\href{https://github.com/mariainesdff/divided_power_algebra_journal/blob/5d1dafcd4c59b7de9791416bfc47442e6b5d2324/DividedPowerAlgebra/DPAlgebra/Init.lean#L291}{\extlink} and \newline
\code{DividedPowerAlgebra.lift_unique}\href{https://github.com/mariainesdff/divided_power_algebra_journal/blob/5d1dafcd4c59b7de9791416bfc47442e6b5d2324/DividedPowerAlgebra/DPAlgebra/Init.lean#L294}{\extlink}.

\paragraph{Divided power structure}

Note that the properties of Definition~\ref{definition:dpa}
mimick the properties~(i), (iv), (v) and~(vi) 
of Definition~\ref{definition:dp}, which is also reflected
by the Lean name~\lstinline{dp} for this function. On the other hand,
properties~(ii), (iii) and~(vii) do not seem to make sense.

However, as its name may suggest, the universal divided power algebra
has a relation with divided powers. This is shown by the following theorem.

\begin{theorem}[Roby, \cite{Roby-1965, BerthelotOgus-1978}]
\label{theorem:roby-dpa-pl}
Let $\Gamma^+_R(M)$ be the ideal of~$\;\Gamma_R(M)$
generated by all elements~$m^{[n]}$, for $m\in M$ and $n>0$.

\begin{enumerate}
\item The map $\iota\colon x\mapsto x^{[1]}$ is an injective linear  map
from~$M$ to its image in~$\Gamma^+_R(M)$;
\item There exists a unique divided power structure~$\gamma$ on
the ideal~$\Gamma^+_R(M)$ such that $\gamma_n(x^{[1]})=x^{[n]}$
for all $x\in M$ and all $n\in\N$;
\item The family $(\Gamma_R(M), \Gamma^+_R(M), \iota, \gamma)$
is universal.
\end{enumerate}
\end{theorem}
Actually, Roby's proof in~\cite{Roby-1965} is slightly inaccurate (in the proof of Lemma~8 in this paper there is an incorrect application of base change of universal divided power algebras),
but the appendix of~\cite{BerthelotOgus-1978} sketches a satisfying argument
along the same lines.
Another proof is given in~\cite{Roby-1968}.

It is not too hard to guess why the proof of part~2 should be delicate.
By definition of the ideal~$\Gamma^+_R(M)$, its elements~$x$
can be written as finite sums of the form 
\[ r x_1^{[n_1]} \cdots x_m^{[n_m]}, \]
where $r\in R$, $m>0$, $x_1,\dots, x_m\in M$, 
and $n_1,\dots,n_m>0$.
If $\gamma$ is a divided power structure on
the ideal~$\Gamma^+_R(M)$, relation~(iv) of Definition~\ref{definition:dp}
applied inductively allows to compute $\gamma_n(x)$
as a finite sum of products of elements
of~$\Gamma^+_R(M)$ of the form $\gamma_i(r x_1^{[n_1]} \cdots x_m^{[n_m]})$,
for which relations~(v) and~(vii) give some expression.
Consequently, there is at most one divided power structure~$\gamma$
on~$\Gamma^+_R(M)$ satisfying the relation~$\gamma_n(x^{[1]})=x^{[n]}$
for all $x\in M$ and all $n\in\N$.
This computation already indicates that
once parts~1 and~2 of the theorem are established,
the proof of part~3 is relatively straightforward.

However, since there are multiple possible choices in writing an element~$x$
of~$\Gamma^+_R(M)$ in this form, and even in applying these relations,
it is not at all obvious that these various choices give rise
to the same value, and even if they would, that the resulting maps
would give rise to a divided power structure.

The rest of this section is devoted to explaining how parts~1 and~2
are established, as well as their formalization in Lean.

\subsection{Base Change}\label{subsec:basech}

 The base change map $m\mapsto 1\otimes m$ induces
 a morphism of $R$-modules $M\to S\otimes_R M$,
 which gives rise to a morphism of $R$-algebras
 $\Gamma_R(M) \to \Gamma_R (S\otimes_R M)$.
 On the other hand, $S\otimes_R M$ is naturally an $S$-module and
 we can consider the divided power algebra $\Gamma_S(S\otimes_R M)$.
 Since there are more relations for $\Gamma_S(S\otimes_R M)$
 than for $\Gamma_R(S\otimes_R M)$, we get a surjective morphism
 of $R$-algebras $\Gamma_R(S\otimes_R M)\to \Gamma_S(S\otimes_R M)$.
 Finally, the composition $\Gamma_R(M) \to \Gamma_R(S\otimes_R M) \to \Gamma_S(S\otimes_R M)$ is a morphism of $R$-algebras that we can enhance into a
 ``base change morphism'' of $S$-algebras
 \[ S\otimes_R \Gamma_R(M) \to \Gamma_S (S\otimes_R M). \]
 This map is in fact an isomorphism, and it satisfies the following uniqueness property.
 
\begin{proposition}[{Roby~\cite[Theorem~III.3]{Roby-1963}}]
\label{proposition:dpa-basechange}
Let $R$ be a commutative ring, let $M$ be an $R$-module
and let $S$ be a commutative $R$-algebra.
There exists a unique ``base change morphism'' of $S$-algebras
\[ \theta_S(M) \colon S\otimes_R \Gamma_R(M) \to \Gamma_S (S\otimes_R M) \]
mapping $s\otimes x^{[n]}$ to $s (1\otimes x)^{[n]}$ 
for every $s\in S$, $x\in M$ and $n\in\N$.
It is an isomorphism of $S$-algebras.
\end{proposition}

In Lean, this proposition is formalized 
as the combination of the \emph{definition} \code{dpScalarExtensionEquiv}\href{https://github.com/mariainesdff/divided_power_algebra_journal/blob/5d1dafcd4c59b7de9791416bfc47442e6b5d2324/DividedPowerAlgebra/DPAlgebra/BaseChange.lean#L177}{\extlink}
of the corresponding linear equivalence,
and a \emph{lemma} that asserts the uniqueness property, called \code{dpScalarExtension_unique}\href{https://github.com/mariainesdff/divided_power_algebra_journal/blob/5d1dafcd4c59b7de9791416bfc47442e6b5d2324/DividedPowerAlgebra/DPAlgebra/BaseChange.lean#L94}{\extlink} (see code excerpt \ref{code:dpa-basechange}).

The existence of the base change morphism is relatively straightforward, but the existence of the
morphism in the other direction is nontrivial. It relies on another universal property\href{https://github.com/mariainesdff/divided_power_algebra_journal/blob/5d1dafcd4c59b7de9791416bfc47442e6b5d2324/DividedPowerAlgebra/DPAlgebra/Exponential.lean#L99}{\extlink} of the universal divided power algebra, using $R$-linear maps from $M$ to the ``exponential module'' of an $R$-algebra $S$.

The same principle allows to prove the following result, whose proof we have formalized\href{https://github.com/mariainesdff/divided_power_algebra_journal/blob/5d1dafcd4c59b7de9791416bfc47442e6b5d2324/DividedPowerAlgebra/DPAlgebra/Exponential.lean#L215}{\extlink}. 

\begin{proposition}[{Roby~\cite[Proposition~IV.8]{Roby-1963}}]
	\label{proposition:roby-quotient}
	Let $P$ be a submodule of~$M$.
	The functorial map $\Gamma_R(M)\to\Gamma_R(M/P)$ is surjective,
	and its kernel is the ideal generated by the elements
	of the form~$p^{[n]}$, for $p\in P$ and $n>0$.
\end{proposition}

\begin{lstlisting}[caption={Base change isomorphism for the universal divided power algebra.},float=htb, label=code:dpa-basechange,captionpos=t,abovecaptionskip=-\medskipamount]

noncomputable def dpScalarExtension :
    S ⊗[R] DividedPowerAlgebra R M →ₐ[S] 
      DividedPowerAlgebra S (S ⊗[R] M) := ...

theorem dpScalarExtension_unique
  {φ : S ⊗[R] DividedPowerAlgebra R M →ₐ[S] DividedPowerAlgebra S (S ⊗[R] M)}
  (hφ : ∀ (n : ℕ) (m : M), φ (1 ⊗ₜ[R] (dp R n m)) = dp R n (1 ⊗ₜ[R] m)) :
    φ = dpScalarExtension R S M := ...

noncomputable def dpScalarExtensionInv  :
    DividedPowerAlgebra S (S ⊗[R] M) →ₐ[S] 
      S ⊗[R] DividedPowerAlgebra R M := ...

noncomputable def dpScalarExtensionEquiv :
    S ⊗[R] DividedPowerAlgebra R M ≃ₐ[S]
      DividedPowerAlgebra S (S ⊗[R] M) :=
  AlgEquiv.ofAlgHom (dpScalarExtension R S M) (dpScalarExtensionInv R S M) ...
\end{lstlisting}

\subsection{The Graduation of the Divided Power Algebra and Polynomial Laws}\label{subsec:dp-gr}

In Definition~\ref{definition:dpa}, the given relations between 
the generators~$x^{[n]}$ are homogeneous, in the sense that
in each relation,
the sum of the exponents that appear in each term is a given integer.
As a consequence, the algebra $\Gamma_R(M)$ has a natural graded
structure for which each element $x^{[n]}$ has degree~$n$, leading
to decompositions as direct sums
\[ \Gamma_R(M) = \bigoplus_{n\in\N} \Gamma_R^n(M), \]
and 
\[ \Gamma_R^+(M) = \bigoplus_{n>0} \Gamma_R^n(M). \]
Moreover, the algebra morphism $\Gamma_R(f)$ deduced by functoriality
from an $R$-linear morphism $f\colon M\to N$ 
is homogeneous, i.e., it maps $\Gamma_R^n(M)$ into $\Gamma_R^n(N)$.

The first two graded terms of $\Gamma_R(M)$ can be
computed explicitly: the canonical map induces
an isomorphism of algebras
\[ R \simeq \Gamma_R^0(M),  \]
while the map $x\mapsto x^{[1]}$ from~$M$ 
gives rise to a linear isomorphism
\[ \iota\colon M \simeq\Gamma_R^1(M) ,\]
proving part~1 of Theorem~\ref{theorem:roby-dpa-pl}.

These isomorphisms are formalized in the definitions\newline \code{DividedPowerAlgebra.ringEquivDegreeZero}\href{https://github.com/mariainesdff/divided_power_algebra_journal/blob/5d1dafcd4c59b7de9791416bfc47442e6b5d2324/DividedPowerAlgebra/DPAlgebra/Graded/GradeZero.lean#L315}{\extlink} and \newline \code{DividedPowerAlgebra.linearEquivDegreeOne}\href{https://github.com/mariainesdff/divided_power_algebra_journal/blob/5d1dafcd4c59b7de9791416bfc47442e6b5d2324/DividedPowerAlgebra/DPAlgebra/Graded/GradeOne.lean#L146}{\extlink}, respectively.

For exponents~$d>1$, the map $m\mapsto m^{[d]}$ 
is not linear anymore, but it is polynomial and homogeneous of degree~$d$.

\begin{proposition}[{Roby~\cite[Theorem~IV.1]{Roby-1963}}]
\label{proposition:roby-dp-pl}
Let $d$ be a natural number.

\begin{enumerate}
\item
The collection of maps
$m \mapsto  \theta_{S,M} ^{-1} (m^{[d]})$ 
from $S\otimes_R M$ to $S\otimes_R \Gamma_R(M)$,
where $S$ ranges over commutative $R$-algebras, gives
rise to an $R$-polynomial law~$\delta_d$
on~$M$, valued in $\Gamma_R^n(M)$.
It is homogeneous of degree~$d$.

\item
The polynomial law~$\delta_d$ is the universal such law: for every polynomial law
$f\colon M\to N$ which is homogeneous of degree~$d$,
there exists a unique $R$-linear morphism $\phi\colon \Gamma_R^d(M)
\to N$
such that $f=\phi\circ\delta_d$.
\end{enumerate}
\end{proposition}
(In this last formula, the linear map~$\phi$ is viewed
as a polynomial law, homogeneous of degree~$1$.)

The Lean formalization of this proposition is still incomplete\href{https://github.com/mariainesdff/divided_power_algebra_journal/blob/5d1dafcd4c59b7de9791416bfc47442e6b5d2324/DividedPowerAlgebra/DPAlgebra/PolynomialLaw.lean}{\extlink}.

The importance of Proposition~\ref{proposition:roby-dp-pl}
comes from the fact
that it informs us about the $R$-module structure
of the graded pieces~$\Gamma_R^n(M)$ of a divided
power algebra~$\Gamma_R(M)$.
Indeed, while its definition makes clear
that the universal divided power algebra is a universal object
in the category of algebras, so that we know how to define
algebra morphisms from a divided power algebra~$\Gamma_R(M)$
to another algebra,
this definition gives no information about
the module structure of~$\Gamma_R(M)$.
This should be compared to the more classical case of polynomial algebras.
Indeed, the definition of a polynomial algebra 
as the free commutative and associative algebra over a module gives its
universal property in the category of algebras,
but gives no information about its module structure,
in particular about its natural grading and
the fact that monic monomials constitute a basis as a module.

Proposition~\ref{proposition:roby-dp-pl}
has the following consequence for the universal divided power algebra of a free module.

\begin{proposition}[{Roby~\cite[Theorem~IV.2]{Roby-1963}}]
\label{proposition:roby-free}
Let $M$ be a \emph{free} $R$-module and let $(m_i)_{i\in I}$ be a basis
of~$M$. For $n \in \N^{(I)}$, write $m^{[n]}=\prod_i m_i^{[n_i]}$.

\begin{enumerate}
\item
The family $(m^{[n]})$,
where $n$ ranges over~$\N^{(I)}$, 
is a basis of the $R$-module $\Gamma_R(M)$.

\item 
For every natural number~$d$,
the subfamily $(m^{[n]})$,
where $n$ ranges over elements of~$\N^{(I)}$ of sum~$d$,
is a basis of the $R$-module~$\Gamma_R^d(M)$.
\end{enumerate}
\end{proposition}

The statement of Proposition \ref{proposition:roby-free} has been formalized in Lean
\href{https://github.com/mariainesdff/divided_power_algebra_journal/blob/5d1dafcd4c59b7de9791416bfc47442e6b5d2324/DividedPowerAlgebra/DPAlgebra/Dpow.lean#L108}{\extlink} \href{https://github.com/mariainesdff/divided_power_algebra_journal/blob/5d1dafcd4c59b7de9791416bfc47442e6b5d2324/DividedPowerAlgebra/DPAlgebra/Dpow.lean#L92}{\extlink},
but not yet its proof.

\subsection{Construction of the Divided Power Structure on the Divided Power Algebra}\label{subsec:dpow-const}
Proving that there can be at most one divided power structure on the universal divided power algebra of a module is easy, since it suffices to check that any two such divided power structures would agree on a generating set of the augmentation ideal (see \code{onDPAlgebra_unique}\href{https://github.com/mariainesdff/divided_power_algebra_journal/blob/5d1dafcd4c59b7de9791416bfc47442e6b5d2324/DividedPowerAlgebra/DPAlgebra/Dpow.lean#L78}{\extlink} for the formalization of this proof). However, proving the existence of this divided power structure is hard, and, as mentioned in~\S\ref{subsec:udpa}, Roby's original proof of existence contained a mistake. More precisely, in the proof of \cite[Lemma 8]{Roby-1965}, Roby misquotes the base change isomorphism for the universal divided power algebra, claiming that there exists an isomorphism
\[ \theta_S(M) \colon S\otimes_R \Gamma_R(M) \to \Gamma_S (M), \]
while the correct claim is that
\[ S\otimes_R \Gamma_R(M) \simeq \Gamma_S (S\otimes_R M), \]
as he had shown in \cite[Theorem~III.3]{Roby-1963} (see Proposition~\ref{proposition:dpa-basechange}).
He then uses the incorrect isomorphism to endow $\Gamma_S (M)$ with a divided power structure induced by the one on $\Gamma_R(M)$. 

However, the main idea of Roby's proof, and also of the proof sketch given in \cite{BerthelotOgus-1978}, consists on reducing the construction of the divided power structure on an $R$-module $M$ to the case of a free $\Z$-module. Our construction of the 
divided power structure on the divided power algebra of a module, which we will now sketch, relies on the same principle.
We have formalized this argument in Lean,
modulo the proof of Proposition~\ref{proposition:roby-free}, which is used as an axiom.
The whole file is quite long, around 900~lines (excluding comments). This length is largely due to the complexity of the definition of the divided power structure on the divided power algebra of a free module\href{https://github.com/mariainesdff/divided_power_algebra_journal/blob/5d1dafcd4c59b7de9791416bfc47442e6b5d2324/DividedPowerAlgebra/DPAlgebra/Dpow.lean#L1066}{\extlink}.

First, we consider the case where $R$ has characteristic zero and recover the divided power structure on $\Gamma_R(M)$\href{https://github.com/mariainesdff/divided_power_algebra_journal/blob/5d1dafcd4c59b7de9791416bfc47442e6b5d2324/DividedPowerAlgebra/DPAlgebra/Dpow.lean#L896}{\extlink} from that on $\Gamma_{\Frac(R)}(\Frac(R)[(X_i)_{i \in \iota}])$, which in turn is given by the fact that $\Frac(R)$ is a $\Q$-algebra.

Next, let $R$ be any ring and let $M$ be a free $R$-module with basis $\mathcal{B} := \{b_i\}_{i \in \iota}$, so that any $x \in M$ can be expressed as a linear combination $x = \sum_{i \in\iota} x_i \cdot b_i$ of the basis vectors. Then the multinomial formula for divided powers\href{https://github.com/mariainesdff/divided_power_algebra_journal/blob/5d1dafcd4c59b7de9791416bfc47442e6b5d2324/DividedPowerAlgebra/ForMathlib/RingTheory/DividedPowers/Basic.lean#L37}{\extlink} yields
a complicated formula for $\gamma_n(x)$ with $n \in \N$ and $x$ in the augmentation ideal of $M$, involving a certain combinatorial coefficient\href{https://github.com/mariainesdff/divided_power_algebra_journal/blob/5d1dafcd4c59b7de9791416bfc47442e6b5d2324/DividedPowerAlgebra/DPAlgebra/Dpow.lean#L448}{\extlink} for which we provide some API. As a consequence, it is hard to verify directly that $\gamma$ is indeed a divided power structure, since this requires to expand concrete formulas and check combinatorial relations.

Instead, we prove that $\gamma$ is a divided power structure by relating it to the divided power structure \href{https://github.com/mariainesdff/divided_power_algebra_journal/blob/5d1dafcd4c59b7de9791416bfc47442e6b5d2324/DividedPowerAlgebra/DPAlgebra/Dpow.lean#L984}{\extlink}
on the free module over the polynomial ring $\Z[(X_r)_{r \in R}]$ with basis a given generating family of $M$.

While this approach allows us to avoid most of the combinatorial computations, it still requires to \emph{define} $\gamma$\href{https://github.com/mariainesdff/divided_power_algebra_journal/blob/5d1dafcd4c59b7de9791416bfc47442e6b5d2324/DividedPowerAlgebra/DPAlgebra/Dpow.lean#L559}{\extlink} using the explicit combinatorial formula, and to check using this definition that $\gamma_n(x)$ belongs to the augmentation ideal of $\Gamma_R(M)$ whenever $x$ does and $n > 0$.
Note that the argument presented in \cite{BerthelotOgus-1978} does not introduce this definition but merely proves its existence. However, from the computational point of view, we found it valuable to provide it.

Finally, we consider the general case of a module $M$ over a ring $R$.

Let $S := \Z[(X_r)_{r \in R}]$ (alternatively, we could let $S$ be any integral domain such that the natural map $\Z \to S$ is injective,
together with a surjective homomorphism  $\phi\colon S\to R$).

Let $F$ be the field of fractions of~$S$.

Let $L$ be the free~$S$-module with basis a given generating family~$(m_i)_{i\in I}$ of~$M$.

To establish the existence of divided powers on the ideal $\Gamma^+_R(M)$
of~$\Gamma_R(M)$, it basically suffices to contemplate the following diagram
of rings, modules and algebras:
\[ \begin{tikzcd}
      F \ar[hookleftarrow]{r}  & S  \ar[twoheadrightarrow]{r} & R  \ar[equal]{r} & R \\
      F \otimes_S L \ar[hookleftarrow]{r} & L  \ar[twoheadrightarrow]{r} & R \otimes_S L \ar[twoheadrightarrow]{r} & M  \\
      \Gamma_F(F \otimes_S L) \ar[hookleftarrow]{r} & \Gamma_S(L) \ar[twoheadrightarrow]{r} & \Gamma_R(R \otimes_S L) \ar[twoheadrightarrow]{r} & \Gamma_R(M)  \\
  \end{tikzcd}
\]

Indeed, the ideal~$\Gamma^+_F(F\otimes_S L)$ has divided powers,
because $F$ is a $\Q$-algebra.

Using the freeness of~$\Gamma_S(L)$ (Proposition~\ref{proposition:roby-free}),
we observe that it is a subalgebra
of~$\Gamma_F(F\otimes_S L)$, and the divided powers of the latter
restrict to a divided power structure on the ideal~$\Gamma^+_S(L)$.

Still using that $\Gamma_S(L)$ and $\Gamma_F(F\otimes_S L)$
are respectively $S$-free and~$F$-free, 
it is not too difficult to deduce,
by quotient, a divided power structure on $\Gamma_R^+(R\otimes_S L)$.

The kernel of the surjective homomorphism
$\Gamma_S(R\otimes_S L)\to\Gamma_R(M)$ is described by
Proposition~\ref{proposition:roby-quotient}, and we can check
that it is stable under its divided powers.  
In the terminology of the theory of divided powers, 
this kernel is a sub-dp-ideal, and
by quotient, we get the desired divided power structure on~$\Gamma_R(M)$.

\section {Remarks about the Implementation}\label{sec:impl}
\subsection{Instances}\label{subsec:instances}
In Lean's core library and in \mathlib, type classes are used to record mathematical structures on types. 
For example, the type class \code{Add} is used to record an addition operation on a type. Here, the \code{class} keyword is used to declare a structure and to enable type class inference for terms of this new type.

\begin{lstlisting}
class Add (α : Type*) where
  add : α → α → α
\end{lstlisting}
Then one can declare \emph{instances} of the class \code{Add} to endow different types with an addition operation (e.g., \code{instAddNat}\href{https://github.com/leanprover/lean4/blob/9d4ad1273f6cea397c3066c2c83062a4410d16bf/src/Init/Prelude.lean#L1719-L1720}{\extlink} is the addition on the type $\N$ of natural numbers). 
Whenever a lemma application requires an instance parameter, Lean will automatically try to synthesize it through a process known as type class resolution.

Type classes can contain both data and propositional fields. For example, the class \code{Ring} contains two operations, addition and
multiplication, together with a list of axioms. Instance parameters are written within square brackets. For example,
\begin{lstlisting}
variable {R : Type*} [Ring R]
\end{lstlisting}
is used to declare a ring instance on the type \code{R}.

Lean's instance resolution procedure will then be able to automatically synthesize related instances, such as the \code{AddCommGroup} instance on \code{R} or the \code{Group} instance on the units of \code{R}.
For a more detailed description of the usage of instances in Lean and Mathlib, see for instance \cite{Baanen-2022}.

In this project, we provide several algebraic instances related to polynomial laws and the universal divided power algebra. In particular, we record as instances 
the $R$-module structure on the type of polynomial laws from an $R$-module $M$ to an $R$-module $N$~\href{https://github.com/leanprover-community/mathlib4/blob/35f7a9bdabaf259310b5434c0e67f4b7b704b96d/Mathlib/RingTheory/PolynomialLaw/Basic.lean#L157}{\extlink}, and the graded algebra structure on the universal divided power algebra \href{https://github.com/mariainesdff/divided_power_algebra_journal/blob/5d1dafcd4c59b7de9791416bfc47442e6b5d2324/DividedPowerAlgebra/DPAlgebra/Graded/Basic.lean#L101}{\extlink}.

Note that global instances should only be used when there is a clearly preferred structure on a given type. For example, as an intermediate step in our construction of the graded algebra structure on $\Gamma_R(M)$, we endow the type \code{MvPolynomial (ℕ × M) R} with a graded algebra structure \href{https://github.com/mariainesdff/divided_power_algebra_journal/blob/5d1dafcd4c59b7de9791416bfc47442e6b5d2324/DividedPowerAlgebra/DPAlgebra/Graded/Basic.lean#L45}{\extlink}; however, we only record this as a local instance, since there are other graded structures one might want to consider on this polynomial ring.

\subsection{Universes}\label{subsec:univ}

In modern set-theoretical language, Russell's well-known paradox shows that there is no set of “all” sets,
and similarly, no formal mathematical definition can ever involve “all” rings, unless the formalism is inconsistent,
so that it is impossible to formalize the definition of a polynomial law exactly as expressed in Definition~\ref{definition:pl}.
This issue is commonplace in category theory and, 
in fact, even if we have given an explicit, relatively elementary definition of a polynomial law, 
this definition can be naturally translated into category theory by saying that it is a morphism of functors.
Indeed, every $R$-module~$M$ gives rise to a base change functor ${\boldsymbol b}_M$ from the category  of $R$-algebras
to the category of $R$-modules, assigning $S$ to $S\otimes_R M$, and up to forgetting the module structure, 
a polynomial law $f\in\mathscr P_R(M;N)$
is nothing but a natural transformation ${\boldsymbol b}_M \Rightarrow {\boldsymbol b}_N$.

Formalizations of category theory have various ways to restrict the mathematical objects
so as to avoid Russell's paradox; a thoughtful discussion is offered by Shulman in~\cite{Shulman-2008}. 
Very often, it is enough to make a distinction between categories and “metacategories”, as is done 
in the classic introductory book~\cite{MacLane-1998}, but then objects such as
morphisms of functors between metacategories would be metametametamorphisms, making them inappropriate
for a formalized study.
In Zermelo--Fraenkel set theory (ZFC), one can consider \emph{universes} 
(introduced by Bourbaki in an appendix to~\cite{GrothendieckVerdier-1972}).
The von Neumann--Bernays--Gödel set theory (NBG) possesses a notion of \emph{class}, a sort of ``large set''
without leading to Russell's paradox, and categories can be classes.
Most type theories have a notion of \emph{universe} which allows to restrict the kind of objects under consideration,
and Russell's paradox is avoided by the fact that the type of all types in some universe does not belong to that same universe.

In our formalization of polynomial laws in Lean,
the rings~\lstinline{S} that appear in the definition of \lstinline{PolynomialLaw} (see code excerpt~\ref{code:pl})
are then restricted to belong to the same \emph{universe}~\lstinline{u}
as the ring~\lstinline{R}. 
In fact, if we wish to be able to take \lstinline{S = R}, this choice is forced by the fact that in
Lean's type theory, an object can only belong to one universe
(“universes  are not cumulative”).

On the other hand, given a polynomial law in this restricted sense,
or, mathematically, for maps $f_S$ as in Definition~\ref{definition:pl}
where $S$ belongs to the same Grothendieck universe as~$R$,
one can deduce analogous maps~$f_S$ for truly all $R$-algebras~$S$,
as well as their compatibility properties.
From a mathematical point of view, this follows from
the fact that any element of a tensor product $S\otimes_R M$
is given by a finite sum $\sum s_j \otimes m_j$,
hence is induced by an element of 
a tensor product $S'\otimes_R M$, where $S'$ is some small $R$-algebra,
e.g., the subalgebra of~$S$ generated by the~$s_j$ (which, being
finitely generated, lifts to the universe~\lstinline{u}).
In our formalization,  we found it easier to take~$S'$
to be a ring of polynomials with coefficients in~$R$ and variables indexed by 
the finite type~\lstinline{Fin n} for some appropriate $n$, which Lean automatically recognizes
as a member of \lstinline{Type u} when \lstinline{R : Type u}.

As a matter of fact, this formalization subtlety is totally overlooked in the initial paper~\cite{Roby-1963},
although polynomial laws are also given an alternative definition which only involves the ring 
$R[T_1,T_2,\dots]$ of polynomials with coefficients in~$R$ in a countable number of indeterminates.

The fact that these maps are well defined, as well as their subsequent
 compatibility properties, can be deduced from
the compatibility properties restricted  to small universes,
the consideration of appropriate diagrams of ring morphisms, and
the \emph{right exactness} of tensor products,
together with the fact that any module is the colimit of its 
finitely generated submodules.

These extensions are formalized in Lean by the declarations
\lstinline{PolynomialLaw.toFun}\href{https://github.com/mariainesdff/divided_power_algebra_journal/blob/5d1dafcd4c59b7de9791416bfc47442e6b5d2324/DividedPowerAlgebra/PolynomialLaw/Basic.lean#L120}{\extlink} and \lstinline{PolynomialLaw.isCompat}\href{https://github.com/mariainesdff/divided_power_algebra_journal/blob/5d1dafcd4c59b7de9791416bfc47442e6b5d2324/DividedPowerAlgebra/PolynomialLaw/Basic.lean#L333}{\extlink}.

\subsection{Rings vs. Semirings}\label{subsec:semir}

The strong tendency to generality in the \mathlib library leads
to the development of many algebraic properties of modules over semirings rather than over rings,
the difference being the omission of the subtraction operator.
Although this level of generality is rarely used in commutative algebra,
our formalization of polynomial laws in Lean
tries to work in this generality, as can be seen from
the use of the type classes \lstinline{CommSemiring}
and \lstinline{AddCommMonoid} in code excerpt \ref{code:pl}.

This raises the question of whether the formalized definition of polynomial law coincides with the mathematical one when the base semiring~$R$
is actually a ring. Indeed, in the \mathlib definition, the maps \lstinline{PolynomialLaw.toFun'}
need to be provided for all commutative semirings which are $R$-algebras, 
while the informal definition only requires these maps for all such commutative rings.
When $R$ is a ring, it contains the element~$-1$ and all commutative semirings
which are $R$-algebras are in fact rings.
Using 
\begin{lstlisting}
let _ : CommRing S := 
  RingHom.commSemiringToCommRing (algebraMap R S)
\end{lstlisting}
we can then tell Lean how to recover the appropriate \lstinline{CommRing}
instance for each $R$-algebra~$S$, in a  compatible way with its initial \lstinline{CommSemiring} instance.
If $S$ already had a commutative ring instance, then the instance obtained using \code{commSemiringToCommRing} will be propositionally equal to the original one, but not definitionally equal, which means that some proofs will require to prove that the two instances agree \href{https://github.com/mariainesdff/divided_power_algebra_journal/blob/5d1dafcd4c59b7de9791416bfc47442e6b5d2324/DividedPowerAlgebra/DPAlgebra/PolynomialLaw.lean#L72}{\extlink}.

We would like to mention that, in order to make it work for semirings,
the formalization of the extension  of 
the definitions \lstinline{PolynomialLaw.toFun'}
and \lstinline{PolynomialLaw.isCompat'} to all universes that we describe in Section \ref{subsec:univ}
required an additional formalization work.
Namely, we had to establish the variant for morphisms of semirings 
of the ring isomorphism $R/\ker(f)\simeq \Range(f)$
when $f\colon R\to S$ is a morphism of rings, 
as well as its two classic companions, and their analogues
for algebras.
In this general context, the \emph{kernel}\href{https://github.com/mariainesdff/divided_power_algebra_journal/blob/5d1dafcd4c59b7de9791416bfc47442e6b5d2324/DividedPowerAlgebra/ForMathlib/RingTheory/Congruence/Hom.lean#L122}{\extlink} of a morphism of semirings
has to be understood as the equivalence relation defined by~$f$,
an object known to \mathlib as a \emph{ring congruence} (\lstinline{RingCon}\href{https://github.com/leanprover-community/mathlib4/blob/f29d34c6e7098de3f46e6323f12ad993c339d2b3/Mathlib/RingTheory/Congruence/Defs.lean#L34}{\extlink})
since it is compatible with addition and multiplication on~$R$.

We formalized the first, second and third isomorphism theorems\footnote{Refer to \cite{wiki:Isomorphism_theorems} for this terminology.} for semirings in \code{RingCon.quotientKerEquivRangeS}~\href{https://github.com/mariainesdff/divided_power_algebra_journal/blob/5d1dafcd4c59b7de9791416bfc47442e6b5d2324/DividedPowerAlgebra/ForMathlib/RingTheory/Congruence/Hom.lean#L340}{\extlink}, \code{RingCon.comapQuotientEquivRangeS}~\href{https://github.com/mariainesdff/divided_power_algebra_journal/blob/5d1dafcd4c59b7de9791416bfc47442e6b5d2324/DividedPowerAlgebra/ForMathlib/RingTheory/Congruence/Hom.lean#L395}{\extlink} and\\ \code{RingCon.quotientQuotientEquivQuotient}~\href{https://github.com/mariainesdff/divided_power_algebra_journal/blob/5d1dafcd4c59b7de9791416bfc47442e6b5d2324/DividedPowerAlgebra/ForMathlib/RingTheory/Congruence/Hom.lean#L400}{\extlink}, respectively. In the same file, we also provided the ring and algebra versions of these isomorphisms.

\subsection{Invisible Mathematics}\label{subsec:invis}

In Andrej Bauer's talk~\cite{Bauer-2023}, \emph{invisible mathematics}
consists in everything that mathematicians skip in their writing,
so that their text is actually legible (Bauer goes up to say that mathematics would otherwise be “unworkable”)
but formally incorrect,
leaving to their readers the task to interpret their statements in the appropriate way.
This can go from implicit introduction of variables, to implicit notation of obvious identifications,
to mere abuse of notation.
Bauer's talk discusses how various techniques in formalized mathematics (such as type classes,
or the use of implicit arguments together with algorithms/tactics to fill them automatically)
allow to make formalized mathematics at all feasible.

In our formalization of the theory of polynomial laws, we 
met these issues at several occasions, most often related to the algebra of tensor products.
 
For example, consider a polynomial law $f\in \mathscr P_R(M;N)$ and
an auxiliary $R$-algebra~$R'$. It happens that $f$ automatically induces
a polynomial law on the $R'$-modules $R'\otimes_R M$ and $R'\otimes_R N$ deduced
by base change, generalizing the classic functoriality in the particular case where  $f$ is a linear map.
By definition, the base change~$f'$ of~$f$ to~$R'$ will consist, for every $R'$-algebra~$S'$,
of a map 
\[ f'_{S'}\colon S' \otimes_{R'} (R'\otimes_R M)\to S' \otimes_{R'} (R'\otimes_R N), \]
together with the natural compatibilities when $S'$ varies, and, expectedly, some relation with
the initial polynomial law~$f$.\\
In fact, to define~$f'_{S'}$, it suffices to consider the diagram

\[ \begin{tikzcd}
S' \otimes_{R'} (R'\otimes_R M) \ar{d} \ar[r, dashed, "f'_{S'}"] &  S' \otimes_{R'} (R'\otimes_R N) \ar{d} \\
(S' \otimes_{R'} R') \otimes_R M  \ar{d} & (S'\otimes_{R'} R') \otimes_R N \ar{d} \\
S' \otimes_R M \ar{r}{f_{S'}} & S'\otimes_R N,
\end{tikzcd} \]
where the first vertical maps are the associativity isomorphisms of tensor products,
and the second ones are the contraction isomorphisms, and then define the dashed arrow $f'_{S'}$ so that this diagram commutes.
Both kinds of vertical maps are typically made invisible in mathematics, 
but the Lean user needs to provide them.

Actually, the second one adds another burden to the code. Indeed, at this point, the data
furnished to the Lean compiler consists of the $R$-algebra $R'$ and the $R'$-algebra~$S'$.
However, in the third line, writing $S'\otimes_R M$ means that $S'$ is viewed as an $R$-algebra,
which is -- obviously to any working mathematician -- done by the composition of 
algebra maps $R\to R' \to S'$.
In \texttt{Mathlib}, an algebra structure is a piece of data
which is provided to the compiler using the \lstinline{Algebra} type class,
and our code has to insert two lines in the middle of the definition of~$f'_{S'}$ that specify this algebra structure \lstinline{algRS'} and 
assert via the type class \lstinline{IsScalarTower} its compatibility with the two other
algebra structures. These lines have to be provided again whenever 
one wishes to do some computations involving the map~$f'_{S'}$,
for example when we prove \lstinline{PolynomialLaw.isCompat'} for~$f'$.
 
\begin{lstlisting}[caption={Declaring an algebra structure within a definition.},float=htb, label=code:pl-algebra,captionpos=t,abovecaptionskip=-\medskipamount]
    letI algRS' : Algebra R S' := RingHom.toAlgebra 
      ((algebraMap R' S').comp (algebraMap R R'))
    haveI istRR'S' : 
        @IsScalarTower R R' S' _ _  algRS'.toSMul :=
      IsScalarTower.of_algebraMap_eq (fun r ↦ by 
        simp [RingHom.algebraMap_toAlgebra])
\end{lstlisting}

This definition of a base change implies that the ground map for~$f'$
is the map $f_{R'}\colon R'\otimes_R M\to R'\otimes_R N$. 
At first we tried to check this compatibility by requesting the computer to identify 
the various objects under display (carelessly, one would say, since the structures
were mathematically equal, but not definitionally so). This
led us to mysterious type class issues, in which the compiler seemed to pick up more or less
arbitrarily one of the possible type classes, so that nothing compiled.
After a bit of thought, we could provide a proof.

Another example of invisible identification of tensor products appears in the study of homogeneity of polynomial laws in $\mathscr P_R(M;N)$, where $M=M_1\times \dots \times M_p$ is a product module.
Indeed, when $S$ is an $R$-algebra, Roby writes~\cite[\S~I.9, p.~227]{Roby-1963} that “one has an obvious 
isomorphism $S[T_1,\dots,T_p] = S \otimes R[T_1,\dots,T_p]$”,
and this isomorphism is never ever written.  In fact, note the \emph{equal} sign used by Roby!
For the formalization, we had to write it explicitely (it is known to \texttt{Mathlib}
as \lstinline{MvPolynomial.algebraTensorAlgEquiv}\href{https://github.com/leanprover-community/mathlib4/blob/91e4c39b71414a092707d1299388397e761f8dd7/Mathlib/RingTheory/TensorProduct/MvPolynomial.lean#L208}{\extlink})
and to establish some additional compatibility properties.

Similarly, we had to provide some API to handle the classical tensor product isomorphism:
\begin{equation}\label{eq:isom}
S \otimes_R (M_1\times \dots \times M_p) \simeq (S\otimes_R M_1) \times \dots \times (S\otimes_R M_p).
\end{equation}
This isomorphism exists in \mathlib under the name\\ \code{TensorProduct.piRight}\href{https://github.com/leanprover-community/mathlib4/blob/cafcdcb2a9130b5471e6ff5fbb0da951cce1d686/Mathlib/LinearAlgebra/TensorProduct/Pi.lean#L86}{\extlink}; however, only a handful of related lemmas are provided, and we found that we needed to add more definitions and results to be able to exploit it.
In particular, we defined three $S$-linear maps: the canonical inclusion of $S \otimes_R M_i$ inside $S \otimes_R (M_1\times \dots \times M_p)$ (denoted \code{TensorProduct.singleRight}\href{https://github.com/mariainesdff/divided_power_algebra_journal/blob/5d1dafcd4c59b7de9791416bfc47442e6b5d2324/DividedPowerAlgebra/ForMathlib/LinearAlgebra/TensorProduct/Pi.lean#L130}{\extlink}), the $i^\text{th} $ projection map from $S \otimes_R (M_1\times \dots \times M_p)$ to $S \otimes_R M_i$ (\code{TensorProduct.projRight}\href{https://github.com/mariainesdff/divided_power_algebra_journal/blob/5d1dafcd4c59b7de9791416bfc47442e6b5d2324/DividedPowerAlgebra/ForMathlib/LinearAlgebra/TensorProduct/Pi.lean#L102}{\extlink}), and the composition \code{TensorProduct.compRight}\href{https://github.com/mariainesdff/divided_power_algebra_journal/blob/5d1dafcd4c59b7de9791416bfc47442e6b5d2324/DividedPowerAlgebra/ForMathlib/LinearAlgebra/TensorProduct/Pi.lean#L166}{\extlink} of these two maps.
By providing an API for these definitions, we get closer to making the isomorphism (\ref{eq:isom}) ``invisible'' in our formalization.

We also provided an analogous API for the special case
\[ S \otimes_R (M_1\times M_2) \simeq (S\otimes_R M_1) \times  (S\otimes_R M_2), \]
which can be found in our file \code{TensorProduct/Prod.lean}\href{https://github.com/mariainesdff/divided_power_algebra_journal/blob/5d1dafcd4c59b7de9791416bfc47442e6b5d2324/DividedPowerAlgebra/ForMathlib/LinearAlgebra/TensorProduct/Prod.lean}{\extlink}.

\subsection{Automation}\label{subsec:auto}
In our formalization of the definition of polynomial law (see code excerpt \ref{code:pl}), we endowed the field \code{isCompat'} with a default proof \code{by aesop}, in the hope that Lean would often be able to prove this compatibility condition automatically. However, so far this has only worked in the definition of the constant polynomial law\href{https://github.com/mariainesdff/divided_power_algebra_journal/blob/5d1dafcd4c59b7de9791416bfc47442e6b5d2324/DividedPowerAlgebra/PolynomialLaw/Basic.lean#L481}{\extlink}, and even in other simple examples (like the polynomial law induced by a linear map) we had to provide this proof by hand.

While working on this project, we found several \mathlib theorems (like \code{Finset.sum_add_distrib}, or \newline \code{TensorProduct.tmul_add}) that had to repeatedly be mentioned in proofs where \code{by simp} would have otherwise succeeded. 
We plan to open PRs to add these results to \mathlib's collection of simp theorems.

Conversely, we also found an instance where simplification was too aggressive and complicated the formalization: namely, the lemma \code{TensorProduct.piRight_apply} simplifies the application of the equivalence \code{TensorProduct.piRight} to its forward direction \code{TensorProduct.piRightHom}, so that one loses the cancellation property with \code{piRight.symm}.

In this case, the definition \code{piRightHom} exists in greater generality than \code{piRight}, so both are needed. Our solution was to exclude \code{piRight_apply} from applications of \code{simp} inside certain proofs (which suggests the possibility of removing it from the \mathlib simp set). Another solution would be to provide a cancellation lemma in terms of \code{piRightHom} and \code{piRight.symm}; however, doing this for general definitions might lead to an undesirable explosion of lemmas, unless these could be automatically generated.

\section {Conclusion and Future Work}\label{sec:concl}
We have formalized the fundamental theory of polynomial laws, providing an extensive API to work with their coefficients and homogeneous components, and setting up the foundations of the theory of divided differentials and divided partial derivatives. 

In parallel, we have formalized the universal divided power algebra $\Gamma_R(M)$ of an $R$-module $M$ and its graded ring structure, as well as a construction of the divided power structure on $\Gamma_R(M)$, modulo the proofs of Theorem \ref{theorem:roby-dpa-pl} and Proposition \ref{proposition:roby-free}, which are left as future work. To prove these results, we will in particular need to prove the Taylor formula for polynomial laws.

Once we complete the formalization of the divided power structure on the universal divided power algebra, we will build over it to formalize the divided power envelope. This is one of the steps in the definition of the Fontaine period ring~$B_\crys$, which we also plan to formalize (note that other main ingredients of its definition, including Witt vector rings \cite{CL21}  and the field of $p$-adic complex numbers \cite{dFF23}, have already been formalized in Lean).

While it is hard to give a precise estimate, the de Bruijn factor for this formalization is quite high. This is due to several factors: to quantify over all $R$-algebras, we had to deal with universes (an issue that is ignored in the pen-and-paper proof but that required approximately 200 lines of Lean code); we generalized the theory from rings to semirings, and we were forced to make explicit many isomorphisms which are treated as (often implicit) equalities in Roby's work. Better infrastructure to deal with instances of ``invisible mathematics'' such as those described in Section \ref{subsec:invis} would significantly simplify the formalization of arithmetic geometry.

\begin{acks}
We are grateful to Kevin Buzzard for suggesting this project and for his support during its completion, to Filippo A. E. Nuccio Mortarino Majno di Capriglio for providing useful feedback on an early version of this manuscript, and to the Mathlib community and reviewers for their advice and insightful reviews on our contributions to the library.

The first author acknowledges support from the \'Emergence project APRAPRAM of Université Paris Cité.
The second author was funded by the Deutsche Forschungsgemeinschaft (DFG, German
	Research Foundation) under Germany's Excellence Strategy -- EXC-2047/1 --
	390685813.
\end{acks}

\bibliographystyle{ACM-Reference-Format}
\bibliography{polynomial_laws_biblio}


\end{document}